\documentclass[letter]{aa}
\usepackage{epsfig}
\def\deg{^{\circ}}                   
\def\etal{{\it et al.}}
\def\ie{{\it i.e.}}
\def\eg{{\it e.g.}}

\def\bei{\begin{itemize}}
\def\eei{\end{itemize}}
\def\bef{\begin{figure}}
\def\eef{\end{figure}}
\def\ben{\begin{enumerate}}
\def\een{\end{enumerate}}
\def\beq{\begin{equation}}
\def\eeq{\end{equation}}
\def\ber{\begin{eqnarray}}
\def\eer{\end{eqnarray}}

\begin{document}
\title{Circulating Subbeam Systems and the Physics of Pulsar Emission}

\author{Joanna M. Rankin\inst{1,3} \and Geoffrey A. E. Wright\inst{2}}
\offprints{Joanna.Rankin@uvm.edu} 
\institute{Sterrenkundig Instituut, University of Amsterdam, Kruislaan 403, Amsterdam 1098 SJ, Netherlands \email{jrankin@astro.uva.nl} \and   
           Astronomy Centre, University of Sussex, Falmer BN1 9QJ  UK 
           \email{gae@pact.cpes.susx.ac.uk} \and 
           On leave from: Physics Deptartment, University of Vermont, 
           Burlington, VT 05405 USA  \email{Joanna.Rankin@uvm.edu}}

\abstract{The purpose of this paper is to suggest how detailed single-pulse
 observations of ``slow'' radio pulsars may be utilized to construct an
 empirical model for their emission. It links the observational 
 synthesis developed in a series of papers by Rankin in the 1980's and 90's 
 to the more recent empirical feedback model of Wright (2003a) by regarding 
 the entire pulsar magnetosphere as a non-steady, non-linear interactive 
 system with a natural built-in delay. It is argued that the enhanced role of 
 the outer gap in such a system indicates an evolutionary link to younger 
 pulsars, in which this region is thought to be highly active, and that pulsar
 magnetospheres should no longer be seen as being ``driven'' by events
 on the neutron star's polar cap, but as having more in common with
 planetary magnetospheres and auroral phenomena.
 
}

\date{Received / Accepted }
\authorrunning{Rankin \& Wright} \titlerunning{Rotating Subbeams \& Pulsar Emission}
\maketitle

%

\section*{Introduction}
\label{intro}

A visitor to a pulsar observing session
will see on the oscillograph something quite unlike anything in the rest
of astrophysics: a never-ending dancing pattern of pulses: sometimes
bright, sometimes faint, sometimes in regular patterns, sometimes
disordered, sometimes switching off entirely only to resurge with
greater vigour. Variations can be found on every time scale down to
tiny fractions of seconds.

Astrophysics is a field used to dealing with objects which evolve over
millions, over thousands of millions of years, perhaps occasionally
punctuated by dramatic cataclysmic events, but generally affording no
more than an unvarying image through the telescope. How are we then to
deal with a phenomenon which is so alien to the common astrophysical experience?

It can be argued that the study of pulsars is more than a
study of complex physics: that it is a study of complexity itself. Beyond
the original insights, some 30 years ago now, that pulsars are
rotating magnetised neutron stars, emitting coherently in the radio
band from a roughly conical region above the magnetic polar caps,
little has been elicited from the welter of information gathered over
the decades to point us towards some fundamental understanding of the
underlying mechanism by which the pulsars emit.

This impasse has arisen partly because pulsars have been
treated primarily as steady-state astrophysical objects undergoing minor
fluctuations which we detect in subpulses, rather than as
intrinsically non-steady, nonlinear systems whose subpulses contain
valuable information about the nature of the system. Yet before any
detailed physics can be undertaken, it is essential to unravel the
embedded complexity and to discern the structure of the underlying
system. This point is well understood in many branches of terrestrial
physics where irregular time series are commonplace. Why is it so
difficult to predict the weather? Why do animal populations
dramatically rise and fall in an apparently random manner? The point
of course is that although complexity may arise through the operation
of complex systems (as with the weather), it can also do so through
simple systems operating under simple conditions---as in the classic
population studies of Prof. Robert May [for a review see May (1986)]. 
And it is essential to distinguish between them, and to know which 
we are dealing with.

In the case of pulsars emphasis has certainly been laid on the former
of these assumptions. Theorists have explored the properties of
time-independent magnetosphere models (often axisymmetric about the
rotation axis, so they would not even pulse!) and assumed that the
observed radio phenomena are complex temporal or geometrical
``perturbations'' of some underlying equilibrium.  Furthermore, many
emission models have seen pulsar ``events'' as being driven and
determined by conditions on the polar cap surface, reflecting the
traditional view of classical dynamics that systems have starting and
ending points, that causality has only one direction.

The problem of this approach is that detailed time-structured
observations have little to say in the construction and verification of 
these models. Perhaps it is possible to take an alternative approach,
well started in a series of papers by one of us and her collaborators 
(``Towards An Empirical Theory of Pulsar Emission'', I--VIII; hereafter 
ETI--ETVIII), to use the observations to determine the model---to ask 
the pulsars themselves how they work.

To do this we will adopt the view that, although apparently complex,
pulsar observations at both radio frequencies and in the optical, x-ray
and $\gamma$-ray regions may be the by-products of a single simple
underlying system. As far as possible special 
pleading or exceptional circumstances will not be introduced in order to explain difficult 
results. The thesis explored in this review is that the simple picture of a dipole 
rotating alone {\it in vacuo}, when inclined at different angles and 
viewed from different angles, can give rise to the myriad of beautiful 
complex phenomena observed in pulsars at many wavelengths over 
the past decades. This thesis will be put to the test.

\section*{Geometry is Pivotal}
Let us assume that the only permanent features of any pulsar are its
underlying magnetic geometry and our particular view of it. Knowledge
of these is the prerequisite to establishing the degree of complexity
(or simplicity) the underlying flow of the emitting particles needs to
possess to account for the highly non-steady observations.

So what results, developed over the many years of pulsar research, can
confidently be regarded as indicators of a pulsar's magnetic field
geometry and thus give a starting point in our quest? Below are listed 
the three most influential ideas, all of which are closely associated 
with a pulsar's most fundamental observational property: its remarkably
stable and individual integrated profile.

\begin{itemize}
\item The most fundamental result---as fundamental today as it was 
over 30 years ago for Radhakrishnan \& Cook (1969; hereafter R\&C) 
and Komesaroff (1970)---is the conal, single-vector-model (SVM) 
geometry implicit in many profile forms and position-angle traverses. 
Without question this is the most successful theoretical idea yet 
articulated as it provides a fundamental standpoint for explaining 
geometric aspects of the observations. Of course, it is probably a 
simplification or abstraction of the actual physical environment.  And 
we must question whether {\it its} underlying assumptions are entirely 
correct.  But (as with the dipolar assumption below) the best means 
of assessing its correctness is to assume it true and then study any 
resulting discrepancies.

\item Second, the extension and development of the foregoing models 
(also Backer 1976) into a profile classification system---the starting 
point of the ``Empirical Theory'' noted above---and their subsequent
evolution into several broadly compatible means of estimating the
magnetic inclination and sightline impact angles $\alpha$ and $\beta$ 
(Lyne \& Manchester 1988; ETVIa,b).  This in turn has led to the 
provisional conclusion that the the integrated emission from most pulsars 
stems from one or all of three different emission beams, the core and 
the inner/outer cones, each roughly centered on the magnetic axis.  

\item Third, it has emerged that pulsar emission beams are nearly
circular!  While various workers have cogently explored whether they
might be latitudinally or longitudinally extended, no strong evidence 
has emerged to the effect that they are non-circular (Biggs 1990; 
McKinnon 1993).  Indeed, probably they are somewhat so, but their 
departures from circularity are evidently small and less systematic than
mere axial extension (Arendt \& Eilek 2003; Eilek \& Arendt 2003).
\end{itemize}

 On the basis of the first two points it may provisionally be concluded
that pulsar emission appears to reflect a magnetic field configuration
which is nearly dipolar in the emission region. While many of us have
at times appealed to ``non-dipolar effects'' to explain sundry
mysteries, no single instance yet exists where this
explanation can be clearly demonstrated. Indeed, although theory of
neutron stars and observations of them in other contexts (\eg, x-ray
binaries) suggest that pulsar surface magnetic fields are probably
{\it not} entirely dipolar---particularly in the case of millisecond
pulsars---our very failure to identify concrete instances of
non-dipolar effects in ordinary pulsars argues that the fields must be
nearly dipolar at the emission-region heights that the observations
reflect.  Furthermore, clear evidence for non-dipolarity will probably
come only by pushing the dipolar assumption so far that
counterexamples emerge. Many theorists have plausibly argued that the
magnetic field in the outer magnetosphere will be distorted by current
flows and relativistic effects (\eg, Michel 1991; Beskin \etal\ 1993; Mestel 
1999; Shibata 1995). But one must be beware of overlooking more
fundamental concepts by using multipole structures close to the
surface to explain difficult observations---\ie, one may fall into the
trap of using complexity to explain complexity.

Support for the third point, also consistent with the dipolehypothesis, 
follows from the identification of circulating subbeams systems in 
B0943+10 (Deshpande \& Rankin 1999) and B0809+74 (van Leeuwen 
\etal\ 2002): it is then this subbeam circulation which produces the 
average conal form, and thus makes them roughly circular in shape---\ie, 
symmetrical about the magnetic axis. The subbeam circulation (identified
observationally as subpulse ``drift'') may be provisionally regarded as 
a general characteristic of conal beams---but the subbeams need not be 
regularly spaced, nor steady over time; they can equally well be formed 
in a sporadic or chaotic manner while still retaining a circular symmetry 
about the magnetic axis.  For these reasons, the form of pulsar beams 
can best be explained by assuming circularity and then assessing any 
evidence for departures.
 
We can therefore adopt three assumptions, the SVM, dipolarity and
conal beam circularity, to jointly provide a standpoint for
constructing simple geometrical models for most pulsars (\eg,
Deshpande \& Rankin 2001; hereafter DR01). To these we can add three
basic electrodynamic concepts, also geometric in nature, which were
established in the early days of pulsar research. First, a light
cylinder, at which corotating particles would attain the speed of
light. Second, a corotating zone whose bounding field line would be
the last to close within the light cylinder; emission would thus be
confined to the open field lines in a region close to the polar cap
and surrounding the magnetic axis. Third, a surface on which the
charge density would be formally zero in a quasi-steady state, and
which would therefore be capable of forming an ``outer gap''
accelerator (Holloway 1975). It is in this last region that $\gamma$-
and x-ray pulses are thought by many (Cheng \etal\ 1986; Romani 1996;
Romani \& Yadigaroglu 1995; Hirotani \& Shibata 1999; Cheng \etal\
2000) to be formed in young pulsars, and it is not unreasonable to
believe that it may continue to play an important role even after its
high-energy phase is past (Chen \& Ruderman 1993; Wright 2003a).

These are the geometric considerations which play a central role in
our approach, but attempts at ``{\it ab initio}'' theorising will be
eschewed: three decades of experience and history have shown that
general pulsar theories---physical theories of pulsars attempting to
deduce the behaviour of real pulsars from first principles---are
incapable of yielding significant, specific, falsifiable expectations
about the observed emission of an actual individual pulsar. Future
more successful theories must be able to do so, and simple
semi-empirical models of the emission geometry along the lines
summarised here provide the essential point of connection between our
natural observations and the ramifications of physical
theories. However in this article, we stress again, the reader will
find geometry put not only to its traditional use of disentangling the
observer's perspective of pulsar ``events'', but given a prominent
role in determining their nature.

\section*{The Pulsar Family}
Although the main focus of this article will be on ``slow'' radio
pulsars, it is important to stress that their properties are likely to
be closely related both to those of faster, younger pulsars such as
the Crab and Vela, which also emit in the high-energy bands, and to the
family of older but rapidly spinning millisecond pulsars.

\subsection*{Young Pulsars}
Through their capacity to produce optical, x-ray and $\gamma$-ray
emission, young pulsars have often been seen as a class apart---not
least because they are observed by a distinctly different community 
of astronomers! Yet this is a dangerous view if we are to regard 
pulsars as exhibiting a continuum of behaviour which evolves as a 
pulsar ages. It has seemed likely that the high-energy photons of 
young pulsars are produced by a different mechanism---and probably 
in a different region of the magnetosphere---from the coherent radio 
emission. It is then easy to believe that those who study radio 
pulsars have little to learn from the high-energy studies, and 
{\it vice versa}.

The stress we are laying on the role of geometric features in
determining phenomena should warn us against this view. Indeed, it is
largely through geometric arguments that the outer gap has been
identified by some (Cheng \etal\ 1986) as a possible source of $\gamma$
rays: and the outer gap is directly linked by magnetic field
lines to what is certainly the site of the radio emission in slower
pulsars. Does outer-gap pair creation cease as soon as the high-energy
emission becomes undetectable? It is possible to construct a viable
emission model in which this process plays a critical role (Wright
2003a), and if verified, could provide a natural link between radio
pulsars and their high-energy siblings.

\subsection*{Millisecond Pulsars}
These pulsars, thought to be older neutron stars which have been
``spun-up'' through a history of accretion, have relatively weak
magnetic fields and often unusual profiles which do not conform to the
pattern of slow pulsars (Kramer \etal\ 1998, 1999). There are good
theoretical arguments for believing that their surface magnetic fields
are highly distorted (\eg, Ruderman \etal\ 1998), which may cause
profile distortion. However, virtually nothing is known of their single
pulse behaviour. For this reason they lie outside much of the analysis
here, but again we would caution against rushing to multipole
geometries as quick explanations. At any large distance from the star 
the dipole component will dominate, and, as we will strongly suggest,
dipole geometries are capable of creating great intrinsic complexity.

\section*{Subbeam Circulation and Pulsar Phenomenology}

\subsection*{Pulsar Profiles as Attractors}
It is no coincidence that the three fundaments listed in the opening
section are all deductions based on the properties of integrated
profiles. A pulsar's profile is its indelible, individual and stable
characteristic. This extraordinary property has been recognized since
the early days of pulsar research.  However, the invariance of
profiles is probably responsible for seducing many theorists into
taking it as evidence of some underlying stability in the emission
system, such that the ever changing behaviour of the individual pulses
can conveniently be ignored.

Yet they are nothing of the sort. Studies of non-linear dynamical
systems have repeatedly revealed the presence of strange attractors,
features which confine the highly time-dependent variables of the
system to a specific region of variable space, but in no way indicate
convergence to a steady state. A pulsar's profile represents a
two-dimensional cross-section (Poincar\'e section) created by our
sightline intersecting an otherwise unseen three-dimensional
attractor. Nothing in the pulsar emits radiation in the form of a
profile. Profiles contain valuable information about the quasi-chaotic
system, but they are not the system itself.

A powerful result of the 1980's was the claim that pulsars have 
attractors in the form of nested cones (ETI, ETVIa), and even that cones
have approximately consistent radii from pulsar to pulsar (relative 
to the size of the polar cap) (ETVIa,b). Over the years there have 
been associated claims that the true attractor structures are less 
(Lyne \& Manchester 1988) or more (Gangadhara \& Gupta 2001, 2003) 
ordered, but nonetheless the implications of these findings remain 
profound. It has long been assumed that pulsar emission emanated 
from particles closely bound to the magnetic field lines, so that the 
emission components followed the contours of that field. The 
consequence of any observations which suggest consistent profile structure
from pulsar to pulsar (such as the ``Empirical Theory'') is then that certain field
lines are preferentially selected by the particles---and very nearly 
the same field lines in each pulsar. Explanations for this in terms 
of the classic Ruderman \& Sutherland (1975; hereafter R\&S) 
model then have to appeal to multipole features in the surface 
magnetic field (Gil \etal\ 2002a,b; Asseo \& Khechinashvili 2002),
yet this begs the obvious question as to why each pulsar would 
have similar multipoles. Alternatively, it has been suggested that 
the cones are formed by multiple refractions within the 
magnetosphere (\eg, Petrova 2000). But then, precisely because 
profiles are only attractors and not the actual emission, we would 
expect the subpulses in the inner and the outer cones to have
similar subpulse behaviour---and this seems to be far from the case.

However, if we abandon the unwritten assumption of these models that
pulsar magnetospheres are systems driven from the polar cap---that the
tiny tail wags the substantial dog---then we are forced to postulate
that somehow the outer magnetosphere {\it selects} the critical
fieldlines. The natural choice for these fieldlines, on both geometric
and physical grounds, would be the cones which connect the outer gap's
upper and lower extrema to the polar cap (as exemplified in the model
of Fitzpatrick \& Mestel 1988a,b). There is anyway strong evidence
that the outer gap plays a critical role in the production of $\gamma$
rays in young pulsars ( Romani \& Yadigaroglu 1995), and it would be
natural that it might continue to play an important, if not directly
detectable, role in slower pulsars.  The opening angles of these
critical field lines seem, on reasonable assumptions about the
emission heights, to have the right proportions to account for the
attractor cones of ET (Gil \etal\ 1993; Wright 2003a), and at these heights the 
magnetic field is almost purely dipolar. It is not impossible that
the precise fieldlines preferred in any given pulsar may be at some intermediate 
value, especially in more inclined pulsars---and may vary in time, 
resulting in multiconal attractors.

\begin{figure}
$$\vbox{
\psfig{figure=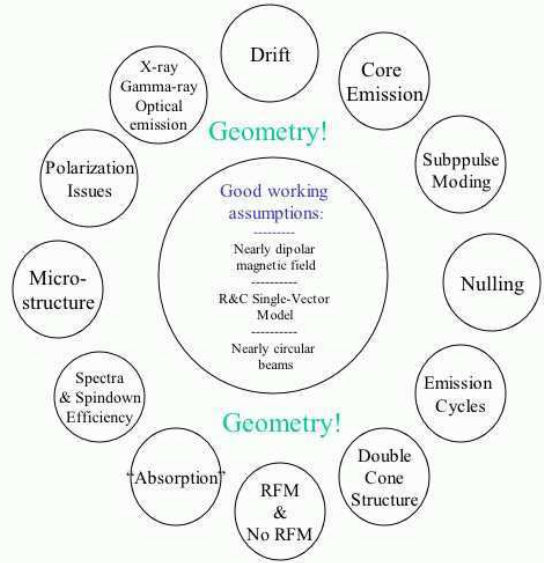,width=8.0truecm,angle=0}
}$$
\caption{A carousel depicting the structure and themes of this paper: 
 the individual topics are linked to the underlying principles via
 their geometric interpretations}
\label{Fig1}
\end{figure}

We are consequently led to understand that it is the {\it downward}-moving 
particles which determine the emission site. These
particles must be accelerated over the vast distances from the outer
gap towards the pole (Mestel 1985; Beskin \etal\ 1993), and particles
of opposite sign must be accelerated back to the gap. This concept
thus shares many features with the free acceleration models of Arons
\& Scharlemann (1979), Mestel (1999), Mestel \& Shibata (1994) and Jessner
\etal\ (2001), although the scale of operation is greater than envisaged
by these authors. More recently, by invoking inverse-Compton
scattering as the principle emission mechanism for producing pairs in
older pulsars, promising models have begun to appear (Hibschmann \&
Arons 2001a,b; Harding \etal\ 2002; Harding \& Muslimov 2002a,b) in
which the acceleration zone is extended further up into the
magnetosphere, and in which pair creation may fail to quench the local
electric field in slow pulsars, thus leaving a residual potential
difference extending to ``infinity''---a feature which could naturally
correspond to the magnetosphere-wide scale requirements of the
empirical model. However, in all these models the implied so-called
``return flow'' should in the present view be seen as the primary
flow, and none have explored the possibility of azimuthally-dependent
emission implied by both observations and the feedback system of
Wright (2003a) (see Figure 2).

The new model may therefore theoretically reproduce the system
attractors---the double cone. But to develop it further on the
empirical basis we have promised above, we must focus our attention on
the pulse-sequence behaviours, and deduce the model's properties from
them. The behaviours can be conveniently discussed under four headings
which summarize four basic emission phenomena: ``drift'', core
emission, mode-changing, and nulling. These headings are largely
suggested by the manner of their detection and observation. However,
it is essential to bear in mind that some or all are often present in
a single pulsar (\eg, B0031--07, B1237+25), and may well spring from
different aspects of the same physical mechanism. A fifth heading,
``emission cycles'', is therefore added, under which we discuss the
apparent ``rules'' or ``memories'' which may link these phenomena. The
principle headings of our discussion are gathered together graphically
in the carousel of Figure 1.

\subsection*{``Drift''/Non-``drift''}
Subpulse ``drift'' is a crucial clue towards solving the pulsar
puzzle, as it exhibits the stunningly beautiful capacity for order in
pulsar radio emission. It is a feature found only in conal regions of
the profile---and indeed only then when our sightline passes obliquely
along the outer edge of the emission cone (thus producing a conal
single, or {\bf S}$_{\rm d}$ profile).  And this drift can range from being
gradual---with subpulses moving slowly across the pulse window over up
to 20 rotation periods---to being rapid---presenting an on-off effect
to the observer. Its intermittent presence in the emission of
predominantly ``slow'' pulsars is powerful evidence of the
unpredictable regularity characterisitic of quasi-chaotic systems. The
emission of some pulsars varies systematically, although not
periodically or even predictably, between discreet drifting patterns
(\eg, B0031--07, B1944+17, or B2319+60), but many/most stars usually
exhibit much less order in their pulse sequences (PSs). No pulsar is
known which permanently emits with one single drifting pattern. On the
other hand, few pulsars have conal emission which is fully
chaotic. Most at least occasionally exhibit sequences which, however
brief, are more or less orderly.

It is possible that higher orders of regularity are present, even in
apparently chaotic emission, which defy detection by current
methods. It may be that we are limited by current analytical tools,
designed to identify specific correlations rather than to measure the
underlying complexity. Power spectra and cross-correlations pick up
strong periodicities at specific phases of the pulse window and
are powerful tools when the emission is highly regular. But how, for
example, could a systematically decaying or oscillating drift rate be
detected?  Near-chaotic systems can exhibit great subtlety in their
behaviour.

How do the differing geometrical circumstances found within the pulsar
population produce the immense variety of patterns---both in the
emission of a single star and among those with ostensibly similar
characteristics?  It is suspected that slow systematic drift over many
periods may be a characteristic of pulsars with small magnetic
inclination angles (well known in this category are B0809+74,
B0031--07 and B0818--13---all thought to be aligned within about
15$\deg$), a result which would suggest that the entire
magnetosphere---and not just conditions near the surface---plays a
role in fixing the subpulse behaviour. However, it is no less
important to understand an unusual ${\bf S}_{\rm d}$ pulsar with no drift,
such as B0628--28, as it is to understand the regularities of B0943+10
or B0809+74, and to account for the more irregular patterns found in
those pulsars with larger magnetic inclinations. Also a puzzle are the
properties of the conal doubles (type ${\bf D}$) stars, where our
sightline cuts the emission cone more centrally (\eg, B0525+21 and 
B1133+16); here some subpulse regularity is observed but apparently
far less than in their close kin, the ${\bf S}_{\rm d}$ stars.

Nonetheless, from both an observational and theoretical standpoint the
natural starting point of any study of ``drift'' is to examine those
pulsars with the most regularly behaved drifting subpulses, and by far
the best and brightest known exemplars are B0943+10 and
B0809+74. Observations of these have given us the telling image of a
circular ``carousel'' of emitting subbeams (Deshpande \& Rankin 1999;
DR01). B0943+10 in particular, when emitting in its highly regular
``B'' mode, exhibits precisely 20 subbeams which circulate around the
magnetic axis about every 37 rotation periods (or about 41 s). This 
star has provided us our first opportunity to {\it count} the number of
subbeams and to confirm the geometric aspects of the R\&S model. Yet
it is now known that even this ``B'' mode adopts slightly varying circulation
speeds on largely unpredictable timespans (Rankin \etal\ 2003). And the
well-known pulsar B0809+74, after being thought for decades to have a
near-clockwork regularity in its drifting pattern, has recently been
found to drift on occasions at a consistently slower rate (van Leeuwen
\etal\ 2002).

The task of accounting for drifting subpulses has only made limited
progress over the years since the publication of the 1975 R\&S polar
gap model. Recently Gil and coworkers have described multipole models
in which ``sparks'' on the polar cap can be made to adequately mimic
the observed drift of certain pulsars (\eg, Gil \& Sendyk 2000), but
this inevitably involves some arbitrariness in the choice of the
magnetic field structure. However, it is possible to produce drifting
subbeams naturally, and without invoking multipoles, through the
operations of the feedback model sketched in the previous subsection (Wright 2003a):
one can suppose the formation of pair-creation ``nodes'' in regions both around
the polar cap close to the surface, and in the outer gap, which ``fire'' particles at each
other and thus create a self-sustaining system.  The nodes will appear
to precess in tandem both about the magnetic axis and around the outer
gap. This system, although still owing much in its physical processes
to the R\&S model (\ie, pair creation and the $\bf{E}$$\times$$\bf{B}$
particle drift), depends on interactions between widely separated
regions of the magnetosphere. Thus a natural time delay is built into
the system, and hence leads to the possibility of chaotic or
quasi-chaotic behaviour. The system can equally well be viewed as
being ``driven'' from the polar cap as from the outer gap, although in
reality it is a self-sustaining system with no starting and no end
point.

\begin{figure}
$$\vbox{
\psfig{figure=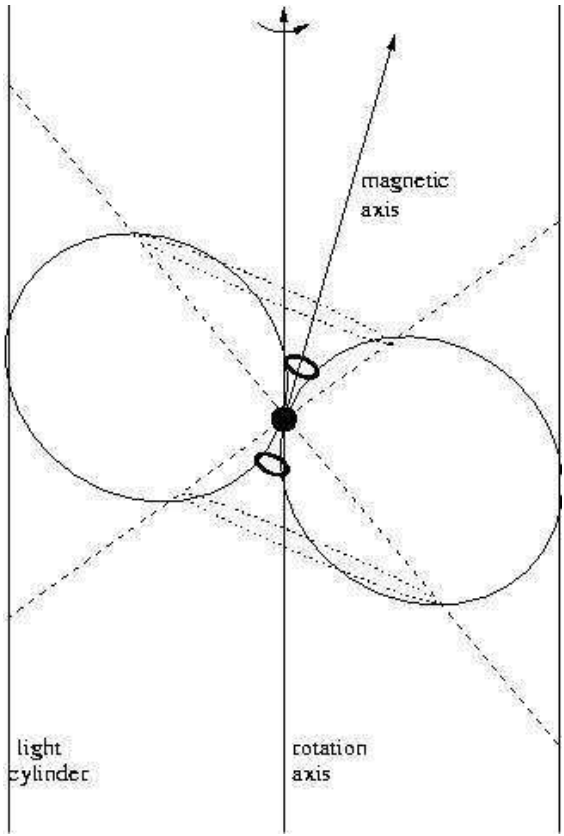,width=8.0truecm,angle=0}
}$$
\caption{The emission geometry for pulsars at a low angle of inclination.
 Note that although the rings of nodes both near the pulsar surface
 (dark ring) and their mirrors on the null surface (finely dotted
 ring) encircle the magnetic axis, only the mirror ring also
 includes the rotation axis. The dotted straight lines representing the null surface separate
 negatively and positively charged regions of the magnetosphere, and
 their intersection with the last closed field-line defines the site
 of an ``outer gap'' [see Wright (2003a) for details].}
\label{Fig2}
\end{figure}

\begin{figure}
$$\vbox{
\psfig{figure=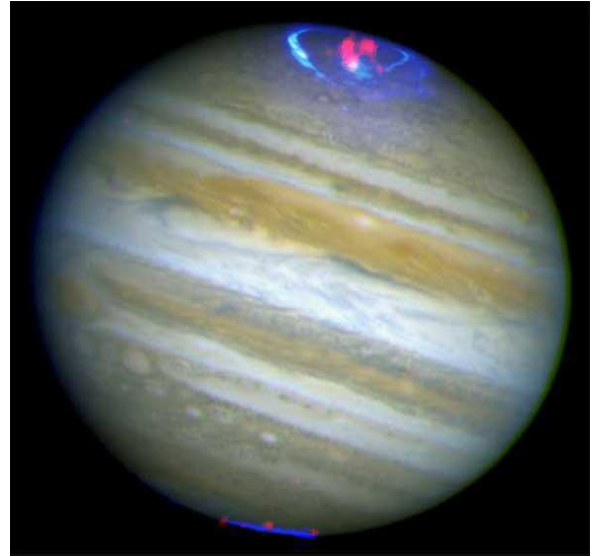,width=8.0truecm,angle=0}
}$$
\caption{A multi-frequency image of Jupiter taken from the Chandra 
 website. The (blue) UV ring and the (red) X-ray spots are superimposed
 on the well-known optical image. It is thought that the X-rays must
 originate from particles reflected from pole to pole. In the movie
 from which this picture is taken the intensity of the spots varies on
 the timescale of the interpole travel time. Image reproduced with
 kind permission from...}
\label{Fig3}
\end{figure}

The promise of this approach is that such a feedback model has within
it the capacity to explain more complex phenomena than the simple
steady circulation of an axisymmetric system. As the magnetic
inclination of a pulsar increases (while yet retaining near dipolar
geometry at relevant heights), the system naturally causes the
emission in the circulating ``carousel'' to develop a patchiness and
asymmetry reminiscent of many observed features. In this view, the
subbeam ``carousel'', although always possessing a near-circular form,
is no more than a distorted ``reflection'' of the outer-gap nodes,
which circulate in tandem with those above the polar cap in an
extended quasi-elliptical path about both the rotation and magnetic
axes (Fig. 2). Although no time dependence is built into it, the model
bears a striking resemblance to auroral models in terrestrial and
planetary magnetospheres [comparisons with the recently discovered
``drifting'' x-ray hot spots around Jupiter's poles (Gladstone \etal\ 
2002---see Figure 3) are particularly apposite], and one may speculate
that phenomena known from these fields---such as flares and magnetic
reconnection---may be found to play an analogous role.

\subsection*{Core Emission}
Core emission, as its name suggests, is emission which appears to be
propagated in a narrow pencil beam surrounding the magnetic axis. Its
angular dimensions are such that, if deemed to be coming directly from
the polar cap surface, it would fill exactly the area enclosed by the
``feet'' of the last closed field lines. A great mystery, of course,
is the relationship between this and the drift emission often found in
the surrounding cones.  We understand the gross distinctions between
them in terms of their beam topology and modulation characteristics
(ETI--V), but we understand virtually nothing about their commonality;
and if the magnetosphere is truly operating as an integrated system,
it seems most likely that both types of radio emission stem from the
same sets of accelerated charged particles. It is tempting---yet at
present no more than a speculation---to see at least a part of the core 
emission simply as the radial reflection of the emission of cascading 
downward flowing particles (Michel 1992; Wright 2003a). Such
particles are an important component of the feedback model, and will
certainly be powerful emitters as they are accelerated immediately
above and towards the polar cap. Above all they will move down the
last closed field lines from the outer gap and naturally define the
limits of the polar cap.

Support for the interdependence of core and conal radiation comes from
pulsars such as B1237+25, where our sightline runs almost directly
over the magnetic axis. In the single pulse trains of this pulsar, the
core region is dormant while the outer components have a strong and
regular periodicity, but when the core brightens (as it does on
quasi-periodical timescales) the conal modulation is interrupted, and
only recommences when the core subsides (Hankins \& Wright 1980). A
case could be made to view the core emission from all pulsars as
generally being inhibitive to regular periodicity in the conal components. 
This is certainly supported by observations of the well known pulsar 
B0329+54 (\eg, Bartel \etal\ 1982; Suleymanova \& Pugachev 1998, 2002;
hereafter SP98, SP02), which has multiple cones as well as a dominant
central core component, yet has never been reported to show any
periodic behaviour.

Surprisingly, core emission has still not been well studied. In part
this is because it was identified after the heyday of enthusiasm for
single-pulse investigations. It is also an unfortunate coincidence
that most of the bright exemplars of core emission lie outside the
declination limits of the Arecibo instrument. This is only a part of
the story, however: the Vela pulsar, perhaps the prime example of core
emission in the sky, has to date been poorly studied. No current 
well measured set of profiles is available, and we do not know
if the star exhibits either polarization or profile modes or if it ever 
nulls.  Many other things have been studied about this nearly unique 
and remarkably influential star (\eg, Krishnamohan \& Downs 1983; 
Radhakrishnan \& Deshpande 2001; Johnston \etal\ 2001; Kramer
\etal\ 2001), but many of the basics remain a matter of guesswork.

The Vela pulsar B0833--45 is probably an excellent example of the 
core-single ${\bf S}_{\rm t}$ class---those with a single core component 
at meter wavelengths.  ETIV has shown that it lies at the short-period 
end of a group whose component widths scale as $P^{-1/2}$---just 
as does the angular width of the polar cap. As the rotation period 
$P$ increases, there is a tendency for stars first to acquire an 
inner cone, and later an outer cone (ETVI). One might therefore
suggest that as pulsars slow and lose their outer-gap high-energy
emission (and by implication their capacity to create self-sustaining
pair production here), sporadic low-energy pair-creation at either 
limit of the outer gap may still be permitted, and this in turn could 
generate conal radio emission through the feedback mechanism outlined 
above (Wright 2003a). As the pulsar further slows, these limits will 
become inaccessible to the sustaining surface x-rays, leading finally 
to the extinction of first the inner and then the outer cone. 

This picture, again based on geometric argument, corresponds well to
the observational analyses of ETI--VI. It also creates, yet again, the
possibilty of a feedback system: when pair-creation becomes prolific
in the outer gap, the downward-moving particles quench the potential
and polar cap pair creation needed for conal radiation (Cheng \etal\
1986). This reduces the heating of the polar cap, which therefore
cools until its thermal x-rays cannot support the outer-gap pair
avalanche, and the mechanism for creating conal radiation can
recommence. Thus, the core emission can be seen as one component of a
thermostatic process!

On the evidence above, the core emission is a very large and
significant piece of the pulsar-emission jigsaw puzzle. In our future
research we therefore should set about answering a series of guideline
questions:

\begin{itemize}
 \item How can we test whether the appearance of a core component in
 the radio emission is evidence of the onset of (possibly short-lived)
 runaway pair-creation in the outer gap? Some kind of statistical test
 for quasi-chaotic behaviour may be appropriate.

 \item Does the core component have significant structure within itself? 
 If the conjecture of Wright (2003a) that core emission is in part reflected 
 conal emission is correct, then the core structure may mimic the conical 
 structure of the outer components. Such structure does seem to occur in 
 B0329+54 (SP02), but in faster pulsars it is often difficult to discern 
 whether the observed structure (Crawford \etal\ 2001) is to be interpreted 
 as truly core or conal.

 \item If core activity is responsible for disrupting quasi-periodic
 conal modulation, can we detect this in the conal emission of pulsars
 which do not have a central sightline traverse? How do the statistics
 of periodicity loss in pulsars with only conal emission compare with
 those where the core is visible? These questions are clearly related
 to the phenomenon of moding, discussed in the next subsection.

 \item How common is quasi-periodicity in core components? And whether
 periodic or not, can any pattern of rise or fall or non-stochastic
 behaviour be discerned?
\end{itemize}

\subsection*{Moding: Changes in Subpulse Patterns}
Historically, this phenomenon has often been associated with, and
identified through, discrete variations in the profile shape. It was
first identified by Backer (1970) in B1237+25, but later in a wide
range of pulsars including B0329+54 (Lyne 1971, SP98, SP02), B1822--09
(Fowler \etal\ 1981), B2319+60 (Wright \& Fowler 1981a), B0943+10
(Suleymanova \& Izvekova 1984; Sulemanova \etal\ 1998; DR01) and most
recently B2303+30 (Rankin \& Wright 2003).  Moding may well be
universal, especially now that even B0809+74---long a considered a
bastion of near-steady regularity (Lyne \& Ashworth 1983)---has been
shown to have a second mode (van Leeuwen \etal\ 2002). This
effectively means that no well studied pulsar has been found to be
free of moding.

However detected, moding is always associated with changes in the
subpulse pattern. In those exemplars listed just above, the moding is
easy to identify through clear and sudden changes in the profile
shape.  In others, such as B0031--07 (Huguenin \etal\ 1970) and
B1944+17 (Deich \etal\ 1986), the mode change is seen as an immediate
and significant change in subpulse drift rate, with later analysis
then revealing an associated profile change (Wright \& Fowler
1982). The changes are often easy to identify, but in some prominent
pulsars exhibiting profile moding without any regular subpulse
modulation (\eg, B0329+54), it is important to identify what changes in
the PSs correlate with the mode changes, work already well started by
Suleymanova \& Pugachev (SP98, SP02).

Interestingly, it seems that at least some pulsars ``anticipate''
their mode changes. This has been demonstrated in both B0329+54
(SP98; SP02) and B0943+10 (Suleymanova \etal\ 1998) where subtle
intensity variations begin some hundreds of pulses before the more
dramatic---almost instantaneous---mode change actually occurs. It is
curious (and hard to account for theoretically) that this slow
anticipatory modulation does not seem, in the case of the exquisite
``drifter'' B0943+10, to affect the periodicity of its drift. Work
is underway to see if B2303+30, which in many ways resembles B0943+10
but whose mode changes are more frequent, also shares this property.

On the basis of these observations it may be useful from a theoretical
standpoint to distinguish between two types of modes: ``ordered''
modes in which the subpulses exhibit regular behaviour such as drift,
and ``disordered'' modes, where the emission is predominantly
chaotic. Thus, B0031--07 and B1944+17 have three modes of the first
kind, B0809+74 (van Leeuwen \etal\ 2002) has at least two of the same,
B1237+25 one of each, and both B0943+10 and B2303+30 may have several
ordered and one disordered mode (Rankin \etal\ 2003; Rankin \& Wright
2003). Often the changes between the {\it ordered} modes may be
gradual, as in B2016+28 (Oster \etal\ 1977).

In most of the pulsars with more than two ordered modes, it has been
found that mode changes do follow some systematic ``cycle'' (generally
accelerating the drift rate through increasing values before returning
to the ``start''). Examples are B0031--07, B1944+17 and B2319+60, and
this may well be a common feature at least of slow-drift pulsars. This
apparent memory is a powerful clue and we must learn how to interpret
it (see the {\it Emission Cycles} section below).

In pulsars with ``disordered'' modes, we must ask if this might always
correspond to the onset of core emission, even if that emission is
fortuitously invisible to us. This question is closely related to our
discussion of B1237+25 in the previous section. The apparently
spontaneous switching from one emission mode to another, whether
ordered or not, is, of course, the hallmark of a quasi-chaotic
system. It need not imply that the switch is ``caused'' by any
external agency either from the interstellar medium or the neutron
star crust. Nor need the moment of change be at all
predictable. Nevertheless, time series generated from these changes
may not be entirely chaotic and it may be possible to borrow
analytical techniques from studies of non-periodic phenomena in other
fields to mine underlying information about the physical system which
produces moding.

Our understanding that the integrated profile forms are produced by a
system of circulating subbeams, which is highly symmetric about the
magnetic axis, constrains our possible interpretations. When a mode
change occurs between ordered modes, it is crucial to know which
parameters have concurrently changed. It was originally believed---for
example in the case of B0031--07---that the repetition rate $P_3$ altered
suddenly, but that the driftband separation, $P_2$, remained
unchanged. This would imply geometrically that the emitting regions,
and their corresponding nodes, would remain on the same fieldlines but
accelerate their drift motion. This appears to be true, at least to
first order, for the five or six pulsars where this phenomenon is
known. Some doubt was cast on this by the discovery that the profiles
of the successive modes did actually widen (Wright \& Fowler 1982), a
result later confirmed by Vivekanand \& Joshi (1997). These latter
authors further claim that their driftband measurements suggest
significant increases in $P_2$ from mode to mode (\ie, as $P_3$
decreases). Such measurements are notoriously difficult to make, not
least because $P_2$ varies across the pulse window and may be subject to
polarization and ``absorption'' effects (see next section). However
van Leeuwen \etal\ (2002) identify a similar effect in B0809+74, where 
an increase in $P_3$ is associated with a narrowing profile and 
an increase in $P_2$.

If confirmed, the change in $P_2$, though slight, would imply (pointed 
out by van Leeuwen \etal\ 2002) that the emission beam has rapidly
moved radially across fieldlines, and in terms of our geometric model
here this would mean the outer gap nodes would have migrated to
different latitudes on the outer gap and the polar nodes to new radii
on the polar cap. This is perfectly possible (given that we do not
know the nature of the underlying cause for the mode change!) and
would reveal interesting properties for the model, but it is first
necessary to confirm this result in more detail: again, it is the
observations which must be the arbiter of the form the model takes.

\subsection*{Nulling}
``Null'' pulses are identified by a complete absence of intensity
throughout the entire pulse window, and appear to interrupt pulse
sequences without warning. Often they persist for many periods (some
pulsars are known which remain in a ``null'' state most of the time),
and then emission reappears as suddenly as it ceased. The phenomenon
is more common in older, longer-period pulsars, although it is no
longer believed that pulsars ``die'' through gradually ``nulling
away'' (see ETIII). In certain pulsars nulls and subpulse drift have
long been understood as closely associated (Unwin \etal\ 1978;
Filippenko \& Radhakrishnan 1982; Lyne \& Ashworth 1983), but we still
have remarkably little physical understanding of these nulls.

It is possible that there are several different kinds of nulls (see
Backer 1971), an idea also hinted at in analyses of the slow-drifting,
moding pulsar B0031--07 (Vivekanand 1995), who found a bimodal
distribution of null length.  Vivekanand did not, however, identify
where the two types of nulls occurred in the PSs. It is very possible
that the shorter nulls tended to occur within a single mode (\ie, the
mode persists following the null) and that the longer ``nulls''
occurred between different modes.  One might also take B0809+74 as
evidence for this idea, as its slower drift mode(s) always seem to
follow long null intervals (van Leeuwen \etal\ 2002).

By contrast, in fast-drifting pulsars there are now strong indications
that nulls are associated with subbeam circulation in a broader
context: pulsar B2303+30 rarely seems to null when in its bright and 
well ordered drift mode, but it exhibits deep nulls in PSs which are
less orderly or chaotic (Rankin \& Wright  2003).  Pulsar B0834+06 exhibits
mostly 1-pulse nulls which appear to fall on the weak phase of its
nearly even-odd PS modulation.  Can it be that in a pulsar (\eg,
B1133+16) with sporadic pulse-to-pulse modulation, there are
occasionally ``empty'' sightline traverses through the average
emission-beam pattern which simply fail to encounter significant
radiation? In order to answer such questions, new investigations of
pulsar nulling are required which investigate the link between nulling
and subpulse behaviour.

A recent result of this new approach is the success in understanding
the null/drifting interaction in B0809+74. van Leeuwen \etal\ (2002, 2003)
have shown not only that each transition to the second mode is
preceded by a null sequence, but that during {\it every} null sequence
the phase of the subpulse is ``remembered'' and then gradually
accelerates either to its previous mode or a new mode. This is a more
subtle interaction than previously suspected (Lyne \& Ashworth 1983).

Knowing the source and growth of nulls in the magnetosphere would
give great insight into their nature. The onset and ending of a ``null''
are so rapid that they are hard to catch in the moment---though such a 
population should occur statistically in many pulsars. We do not know
how close to simultaneous is the onset at different frequencies (and
by implication at differing locations in the emission zone), nor
whether it ends as fast as it commences.  Currently the Multi-Frequency,
Multi-Observatory Pulsar Polarimetry (MFO) Project (Ramachandran 2002) 
is gathering simultaneous broad-band observations which are providing 
the first general opportunity to address such questions. Does the entire 
``carousel'' of subbeams switch off together? A study of nulling in 
conal double ({\bf D}) pulsars might help resolve this. From an 
observational point of view it is difficult to distinguish between short 
``nulls'' (with absolutely no emission) and very weak emission, so the 
observations required are not easy to obtain.

The statistics of null and burst length in a given pulsar are as
important as the null fraction. {\it Where} null pulses occur within
PSs is equally crucial. An interesting analogy to nulling---and to
mode changing too---is the incidence of terrestrial earthquakes or
avalanches, where larger earthquakes (avalanches) occur less
frequently according to a specific law (the Gutenberg-Richter law). It
is known that statistics of these ``self-ordered critical systems'' (SOC's) 
reveal characteristics of the underlying physical processes, 
typically through the presence of power-law distributions (Bak
1996). This procedure has also successfully been applied to solar
flares and magnetic reconnection, and would be interesting to pursue
in this context.

There are virtually no working theories which adequately account for
nulling, and hence there is no agreement as to whether nulling occurs
because the engine producing the emitting particles temporarily
``switches off'', or whether the emission process itself breaks
down. For example, in the feedback model outlined here, the flow from
the suface to the null line and back may not be continuous and may
contain irregular or even ``void'' stretches of low particle density
creating voids in emission, which we experience as
nulls. Alternatively, or additionally, nulls may arise through a
breakdown in the mechanism which maintains the coherent
radiation. This latter could arise simply because the rapidly changing
flow cannot hold the flow steady enough for the conditions producing 
coherence to develop. Thus the nulling phenomenon could be seen 
as the visible yet `superficial' response to one of a range of deeper 
underlying conditions. One might then predict that nulls will be more 
prevalent in highly irregular stretches of emission, and there is some
suggestion of this in the observations of a number of pulsars (\eg,
B2303+30), but it is important to test this in more careful analyses
of observations.

\subsection*{Emission Cycles}
In a number of pulsars, so far 4, the emission modes are characterized
by a progressive increase in drift-rate through at least 3 modes. These 
pulsars [B0031--07 (Huguenin \etal\ 1970), B1918+19
(Hankins \& Wolszczan 1987), B1944+17 (Deich \etal\ 1986), B2319+60
(Wright \& Fowler 1981a)] all have very low drift-rates in their
principal modes, and the magnetic axes of all are thought to be weakly
inclined with respect to their rotation axes.

What is remarkable about them is that the mode sequences appear to
follow certain ``rules''. For example, B0031--07 has 3 identified modes,
A, B and C, which have repetition periodicities ($P_3$s) of about 12, 8
and 5 periods respectively. These modes are interspersed by null
stretches, both within a mode and between modes. But often a
mode-change occurs without an intervening null, and then it is found
that only transitions A to B or B to C are allowed (Huguenin \etal\ 1970; 
Wright \& Fowler 1981b). The transitions take place within a
few rotation periods at most, possibly within a single period. In
other words, sudden, null-free mode changes in which the driftbands
retain their identity can only occur when the mode change corresponds
to an {\it increase} in the drift rate. This gives the impression that
the pulsar emission is executing a kind of cycle, from A to B to C,
which may last some hundreds of pulses. Not all cycles include C,
which is anyway of short duration.

These properties are shared by the remaining 3 pulsars, and recently
the pulsar B0809+74, also with a slow drift rate and low inclination,
has also been shown to have smooth transitions (preserving drift-band
identity) from a slow mode to fast, but only fast to slow following a
long null (van Leeuwen \etal\ 2002). All this suggests that the
subpulse sequences possess direction, even ``memory''. This behaviour
is known in many branches of non-linear studies. Near-chaotic systems
can move from one pattern to another, apparently unpredictably,
migrating from one limit cycle to another with a corresponding change
in attractor. Regarding profiles as attractors, this is precisely what
the changing pulsar profiles reveal: the profiles of the emission modes
do seem to widen as the driftrate increases, and by a similar amount
in each pulsar.

A further curious fact about these pulsars is that the ratios of their
successive modal drift rates are about the same: all (including B0809+74)
appear to increase their driftrates by about 1.6 as they move from one
mode to the next. Whether this increase entirely stems from a
reduction in the pattern repetition rate ($P_3$), or whether also the
band spacings ($P_2$) are slightly reduced [as Vivekanand \& Joshi (1997)
and van Leeuwen \etal\ 2002 find for B0031--07 and B0809+74
respectively] is important to clarify.

It is fascinating to speculate as to what is physically happening
during these ``cycles''. It seems unlikely that additional nodes
(Wright 2003a) or sparks (Gil \etal\ 2002a,b; Gil \& Melikidze 2002) are created as the
mode transitions occur, for they seem to be smooth and no act of
node/spark creation is observed. The conclusion is that the emission
region, and hence the nodes/sparks, must migrate radially to an inner
set of field lines. In Wright's model, the outergap mirror points must
move along the gap further from the star. The cycle would then begin
with a slow drift in outer fieldlines, possibly those bounding the
corotating zone, and progress towards the axis. This spiral inwards is
reminiscent of the model suggested many years ago for B1237+25 by
Hankins \& Wright (1980), although in this strongly inclined pulsar
the entire sequence lasts only 2.8 periods. Note also that the slow
drift rate of the weakly-inclined pulsars implies (in the model of Wright 2003a) that the outergap
nodes are nearly corotating with the star---appropriate to the
corotating zone boundary. Then, as the modes progress, they spin
faster ({\it in the corotating frame}) counter to the rotation of the
star, and eventually become closer to being stationary in the observer
frame and nearer the light cylinder.

The fact that we have only 4 established examples so far of this
cyclic behaviour may be because lengthy and detailed studies of the
subpulse sequences is necessary before the mode-change ``rules'' 
become apparent. But if subpulse behaviour does result from 
magnetospheric feedback (Wright 2003a), it also may be because at 
the more common larger angles of inclination (say between 20$\deg$ 
and 50$\deg$) the structure of the magnetosphere's potential becomes 
highly asymmetric, with both faster driftrates and more blurred mode 
changes. Hence slow long-term cycles may only be a feature of 
nearly-aligned pulsars.

\section*{Integrated Profile Questions {\it or} Attractor Analyses}
Although we have stressed the importance of single pulse analyses and
implied that they are the true currency of a pulsar's emission, the
fact remains that single pulse analysis is only possible for a small
minority of the pulsar population. Only 10--20 pulsars have so far
had their single pulse behaviour documented, and for some of these the
description is only preliminary [\eg, B1112+50 (Wright \etal\ 1985)] or
relatively inaccessible (\eg, Ashworth 1982; Backer 1971). Many more
bright pulsars, some recently discovered (\eg, Lorimer \& McLaughlin
2003), are deserving of greater study and we feel a major effort
should be made to accelerate their analysis.

We are therefore forced to accept that for the majority of pulsars
(there are some 1700 known to date) only their integrated profile is
available for study. Yet, so long as we continue to bear in mind their
nature as attractors, it is possible to mine a great deal of information, 
particularly of a geometric nature, from this population. This is 
essentially what was done by one of us (ETI,ETII,ETIII) and Lyne \& 
Manchester (1988) in the eighties, when perhaps 200 usable profiles 
were available.

What now needs to be done is to look at these results in the light of
our twin hypotheses of nearly circular subbeam circulation, and a
pan-magnetosphere feedback system. We will take in turn a number of
crucial aspects of profile analyses, and pursue the consequences.

\subsection*{Double-Cone Beam Structure} 
While several pulsars with five distinct components had long been 
known (\eg, B1237+25, B1857--26), suggesting two conal rings as well as 
a central core beam, it was not until 1993 (ETVI) that firm evidence 
was given for two cones with coherent geometrical characteristics (then
confirmed by Gil \etal\ 1993; Kramer \etal\ 1994). 
Specifically, the respective inner and outer cones in double-cone 
({\bf M}) stars were found to have outside, half-power, 1-GHz radii 
of 4.33$\deg P^{-1/2}$ and 5.75$\deg P^{-1/2}$---and the single cones 
of triple ({\bf T}) pulsars were found to be one or the other.
Then, ETVII showed that while outer cones exhibit RFM, inner-cone 
radii appear to be nearly constant over the entire radio band.  

It is still not understood why some pulsars have two concentric cones
and what is the relation of the subpulse modulation in the two cones.
Even for the paragon of this phenomenon, B1237+25, there is very much
still to learn.  One line of approach has assumed that double-cone
emission probably comes from the same set of emitted charges, whether
sparks or beams (\eg, Deshpande \& Rankin 2001), while Gil and
collaborators (Gil \& Sendyk 2000; Gil \etal\ 2002a,b; Gil \& Melikidze 2002) have
envisioned several concentric rings of sparks on the stellar surface
which are thought to be associated with the various cones and even the
core.  The implications of these assumptions are very different, and
almost no work has been devoted to pursuing their study through PS
analyses.  There are now almost 20 stars with well identified 
double-cone profiles (see ETVI), so a systematic study is possible and
feasible---though perhaps no more than half a dozen are strong enough
for PS analysis.  In any case, that rotating subbeams are responsible
for the generation and modulation of these cones gives us new ways of
studying and assessing their character and origin.

\subsection*{RFM/no RFM}
The phenomenon that prominent conal double profiles (\eg, B1133+16) 
become progressively wider with wavelength was well noted very early 
(Komesaroff 1970), and many workers participated in documenting the 
effect, often by fitting pairs of power-law functions to the asymptotic 
high and low frequency profile widths or component spacings (\eg, Lyne 
\etal\ 1971; Sieber \etal\ 1975).  Thorsett (1991) demonstrated that a
function of the form $\varphi_0$+$(f/f_0)^{-a}$ fitted the full low 
to high frequency behaviour well.  von Hoensbroech \& Xilouris (1997) 
have provided a full review of work on ``radius-to-frequency mapping'' 
(hereafter RFM) in the course of extending the range of the high 
frequency observations.

The question of RFM has perhaps become more interesting with the 
conclusion that it is exclusively a characteristic of outer-conal 
beams (Mitra \& Rankin 2002; ETVII).  This study included the beam 
geometry for the first time, so that conclusions are framed in terms 
of conal beam radii and emission heights.  In addition, two different 
types of RFM behaviour were identified among the outer cones---those 
which approach a constant radius at the highest frequencies and those 
that do not.  

Nonetheless, in many other ways the emission from inner and outer
conal beams is indistiguishable, so that it again becomes a question
of how a rotating subbeam system radiates in such a manner that its
envelope does or does not exhibit a frequency dependent radius.
Closely related questions arise in considering the significance of the
altered profile forms produced by mode changes.  This was first noted 
in the context of pulsar B0329+54 (Lyne 1971), but excellent examples 
are now also B0611+22 (Nowakowski \& Rivera 2000) and recent 
studies of B0809+74 (van Leeuwen \etal\ 2002, 2003).

The problem is aggravated by the fact that there is no agreed model
for the production of emission.  Assuming that the coherent radiation
is emitted tangentially to field lines in the polar cap region some
hundreds of kilometers above the stellar surface, the critical problem is
then to determine precisely which fieldlines are carrying the emission
and at which height (Kijak \& Gil 2003). This is no easy task given the likely effects of
aberration and time-delay (Gangadhara \& Gupta 2001; ETVII). From the
standpoint of the feedback model, this work is very important, since
it will determine the nature and true positioning of the link to the
outer gap.

\subsection*{``Absorption''}
This phenomenon relates to the gross asymmetry which is evidenced in
certain pulsar profiles, yet only within certain frequency ranges,
suggesting that the asymmetry is not an intrinsic property of the
profile but that some intervening medium has partly ``absorbed'' the
emission. Although first discussed in the context of multi-frequency
alignment anomalies in the relatively stable pulsar B0809+74 (Davies
\etal\ 1984; Bartel \etal\ 1981; Bartel 1981), where the drifting subpulses
become blurred and attenuated as they pass through a specific longitude. 
range.  The phenomenon is now also known to be closely associated 
with profile mode changing, as (\eg, in B0943+10) the degree and 
character of the ``absorption'' is strongly correlated with the profile 
mode.  Thus much of what was said above in regard to profile modes 
is equally applicable here.  Indeed, perhaps profile-mode changing 
and ``absorption'' should be viewed together as two faces of one
phenomenon---the temporal and profile-spectral manifestations of a
single cause which is also manifested in the PS pattern. From an
observational standpoint we now can see why ``absorption'' is most
clearly or usually identified in pulsars with $|\beta|$/$\rho$ near 
unity (\ie, usually stars which are members of the ${\bf S}_{\rm d}$ 
profile class) whereas profile mode changing is most easily identified 
in multiply-peaked pulsars with $|\beta|$/$\rho$ much less than 
unity.

A more recent variation of this topic has come with the discovery of
mysterious ``notches'' in the profiles of a number of very different
pulsars (McLaughlin \& Rankin 2003). These are found in certain
wide-profile pulsars, are narrow double-dip in character, and tend to
follow the profile centroid by about 60$\deg$ longitude. The critical
question is whether they are intrinsic to the profile or are truly
absorption. There is a need for frequency-dependent studies to 
resolve this, since if not intrinsic the notches could for the first time
identify very localised regions of the magnetosphere where the
absorption takes place.

Defining the location of absorption, given the geometric theme of this
paper, is something of a challenge. Assuming that the effect is not
occuring within the circulating subbeams, we have to track the likely
path of the radiation as it escapes the magnetosphere. Three things
will be crucial to this: 1) the longitude in the profile, which defines 
the moment and angle of emission, 2) the rotation period, which 
fixes the scale of the magnetosphere, and 3) the angle of
inclination, which locates the position of the null surface, the outer
gaps and the distance from the magnetic pole to the light cylinder.

A start on this problem very much in keeping with the ``keep-it-simple''
geometrical ideas of this article has recently been made on the issue
of profile notches: it can be shown that if, as generally assumed, the
emission frequency is height-dependent, then double-notches can arise
through time-delay and aberration in quite natural geometries and
needs no appeal to ``distorted'' field-lines (Wright 2003b).

\subsection*{RF Spectra, Rotation Energy Loss \& RF Efficiency} 
That conal beams are produced by systems of rotating subbeams gives 
us the possibility of estimating the full radiation pattern of a given pulsar 
in terms of what we observe in the course of our particular sightline 
traverse.  A first effort in this direction was made by Deshpande \& 
Rankin (1999) for pulsar B0943+10.  Then, the emission from the full 
beam pattern at a given frequency can be integrated over the pulsar's 
full spectrum and compared with its rotational energy loss to estimate 
its overall RF radiation efficiency.  Such efforts, carried out for a 
substantial group of stars promise to provide an important quantitative 
point of connection with physical theories of pulsar emission.  

The above work depended on low frequency observations made over 
many years at the Pushchino Radio Astronomy Observatory and the 
catalogues of pulsar spectra and luminosity estimates compiled by 
Malofeev and colleagues there (\eg, 1996, 1999).  Otherwise, only 
limited progress has been made in understanding why pulsars exhibit 
different radio-frequency spectra.  Some attention has been paid  
to spectral-index differences at centimeter wavelengths (\ie, Maron 
\etal\ 2000) as well as the breaks in such indices exhibited by certain
stars (Ochelkov \& Usov 1984; Beskin \etal\ 1988; ).  However, equally 
important is the issue whether a pulsar's spectrum turns over at low 
frequencies (Benford \& Bushauer 1977; Malofeev \& Malov 1981).  
Some pulsars (\eg, B0329+54) exhibit spectral turnovers at 100--300 
MHz, and thus are observable at low frequencies only with great 
difficulty, if at all. Other pulsars---and it seems all of those best known 
for their regular drifting-subpulse systems (\eg, B0031--07, B0809+74,
B0943+10)---are observable to very low frequencies.  B0943+10 in
particular exhibits no spectral turnover down to some 30 MHz 
(Deshpande \& Rankin 2001).  It should thus now be possible to gain 
some insights into the physical reasons for such different behaviours.

\section*{Emission Questions}

\subsection*{Microstructure and Giant Pulses} 
A number of observations and developments have begun to narrow the
possible interpretations of microstructure.  Apparently, the fine temporal 
structure of microstructure has been resolved, and additionally there is evidence 
that its autocorrelation length scale is roughly proportional to the pulsar 
period (Popov \etal\ 2002).  Observations using the Effelsberg and 
Westerbork instruments seem to show in pulsars like B0329+54 that 
micropulses occur in all three main components (Lange \etal\ 1998; 
Ramachandran 2001) so that the phenomenon---as with nulling---affects 
both core and conal components. It is, however, far from clear what 
connection, if any, micropulses have with subbeams in {\bf S}$_{\rm d}$ 
stars such as B0809+74 or whether there is any orderliness to their 
polarization characteristics.  Much work then needs to be done in order 
to assess how closely associated microstructure is with the other primary 
pulsar phenomena---and the study of microstructure in a pulsar with a 
very orderly rotating subbeam system such as B0809+74 undoubtedly 
has much to teach us about the nature of microstructure.

In a small, but now growing, number of fast pulsars microstructure is
found to be associated with the phenomenon of ``giant pulses''. Such
pulses are narrow but exhibit an intensity far in excess of the mean
pulse energy. They were first found in the Crab pulsar (Heiles \etal\
1970; Lundgren \etal\ 1995; Hankins \& Kern 2003), but recently have 
also be found inspun-up millisecond pulsars (\eg, Kinkhabwala \& 
Thorsett 2000). Theyhave a distinctive power-law energy distribution---possibly
suggestive of self-organised criticality (see earlier section on {\it
Nulling}, and Young \& Kenny, 1996), suggesting they might be the
response to a simple generating physical system operating on a wide
range of scales in a self-similar manner. Furthermore, they often
occur at phases close to those of the high-energy profile components
(Romani \& Johnston 2001). Assuming high-energy emission is indeed
emitted from outer gaps, all this intriguingly hints at an independent
physical radio source for giant pulses and at an interrelationship
between polar cap and outer gap {\it radio} emissions.

\subsection*{Polarization Issues} 
The origin of the orthogonal polarization modes (OPM) is one of the
great mysteries of the pulsar emission problem.  Those stars so far
well studied in terms of their OPM characteristics are so far almost all
conal dominated, so we have virtually no good examples apart from 
B0329+54 (Bartel \etal\ 1982; Gil \& Lyne 1995; SP98, 02) of core 
components that can indicate what role the OPM play in core 
emission.  What is increasingly clear is that conal pulsar beams are
highly modal in their angular beaming characteristics (Deshpande \&
Rankin 2001; Rankin \& Ramachandran 2003; Rankin \etal\ 2003).  
OPM has historically been assumed to be a characteristic of the 
pulsar emission mechanism, but there is now theoretical work to the 
effect that it may result from propagation effects (Arons \& Barnard 
1986; Petrova 2000). Basic questions about whether the two modes 
occur simultaneously in individual samples and whether they are fully 
or partially polarized remain.  The issue of how such characteristics
vary with frequency, and whether they might be implicated in the
general lower levels of polarization at very high frequencies, has 
only been touched.  Again, each of these questions can fruitfully be
studied in the context of a rotating subbeam system, because the
modulated character of the signals gives one an additional method by
which to separate the combined effects of the modal interactions. The
work of McKinnon \& Stinebring (1998, 2000) has provided a much
sounder statistical and interpretive foundation for OPM work, but
analyses based on sensitive and fully calibrated recent observations
are needed to carry this work further.

Just why pulsars have polarization and depolarization remains a mystery.  
The modal character of the rotating subbeam systems which produce 
conal beams is probably almost entirely responsible for the complex 
variety of characteristics observed in different pulsars with conal profiles.
Rankin \& Ramachandran (2003) have explored the character of this
beam-edge depolarization in stars with conal components pairs where
our sight-line passes close to the magnetic axis as well as in conal
single (${\bf S}_{\rm d}$) stars where our sightline makes an oblique
traverse, finding that a virtually identical beam configuration can
produce the full range of observed effects.  Therefore, it should be
possible to model the depolarization of the conal emission in a wide
variety of situations to both improve our knowledge of the conal
emission geometry as well as the nature of the modal emission which
produces the polarization and depolarization.  This is a rich area for
immediate study and a good example of how the context for PS analysis
and interpretation has been changed almost completely by our
expectation that rotating subbeam systems produce the emission which
both polarizes and depolarizes conal components.

A closely related question is the polarimetric relation between inner
and outer conal beams.  We often see evidence for two active modes on
the outer edge of the emission beam and only one in interior regions
of the profile (or beams).  In a double-cone star (\eg, B1237+25) this
means both modes are active in the outer cone, but only a single mode
is apparent in the inner cone.  However, we also see cases of
inner-cone {\bf D} or {\bf T} stars where both modes are also active
in their inner cones.  Must this circumstance not bear importantly on
how the OPM is generated---that is, whether it is an emission or
propagation effect?

\subsection*{X-ray, Optical \& $\gamma$-ray Emission}
We close our carousel of pulsar phenomenology (Fig. 1) with a
discussion of high-energy pulsar observations and their relation to
the radio emission.

Although nearly 2000 radio pulsars have now been discovered, only a
handful of these have been detected at high-energy wavelengths. Future
satellites promise to greatly expand this number, but there remains a
feeling that we are dealing with two classes of pulsars: one
population where the x-ray and shorter wavelength emission is closely
correlated with young pulsars having large values of $B_{12}/P^2$
together with prominent or exclusive core emission beams at radio
wavelengths, the other with no high-energy emission and exhibiting
predominantly conal features. The contrast is exacerbated by the fact
that pulse-sequence analysis is still impossible for the high-energy
emission, and that therefore only profile (\ie, attractor) studies are
available. This has further fostered the impression that such pulsars
are in a steady state, and furthermore emit from different regions by
a different mechanism.

Yet this dichotomy may be an illusion, brought about simply by the
differing means by which the high-energy and radio emisson are
detected: a major point stressed in Wright's (2003a) work is that even
in slow pulsars the radio emisson may require the interactions between 
differing regions of the magnetosphere. Thus a highly-energetic pulsar
with core emission and an old pulsar with only outer cone radio
emission above the level of detectability represent opposite ends of
an evolutionary spectrum. Younger pulsars with energetic outer gaps
would ``quench'' the electric field in the conal regions above the
poles (Cheng \etal\ 1986), and the {\it downflowing} particles would
somehow generate the core emission, possibly by reflection of incident
radiation generated just above the polar cap surface. 

Measurements of correlations between core and high-energy emission
must become possible in the near future and provide an opportunity to
test and develop these ideas. For example, certain x-ray pulsars, such
as B0611+22 (Nowakowski \& Rivera 2000), exhibit slow quasi-periodic
profile changes which could be interpreted as a very slow rotation of
one or more subbeams (Kern 1998), and it would be interesting to know
whether the x-ray emission is modulated or correlated with these radio
profile variations.

The evolutionary picture has received some theoretical support from
the recent studies by Harding and her coworkers (Harding \etal\ 2002;
Harding \& Muslimov 2002a,b) of pair creation in the polar cap region.
Building on the earlier work of Hibschman \& Arons (2001a,b) which
incorporates the effects of backflowing particles in determining the
height of the acceleration zone, they envisage a pulsar's radio
emission as a two-phase process: above the polar caps of young pulsars
the principle radiation mechanism by which pairs are produced is
curvature radiation, which generates sufficient pairs to screen the
ambient electric field. However in older, ``slow'' pulsars, this gives
way to radiation dominated by inverse Compton scattering---and
crucially the electric field above the acceleration zone cannot now be
fully screened. This implies that particles will continue to be slowly
accelerated towards the outer gap, where Wright (2003a) envisages the
occurence of further pair creation. Intriguingly, Hirotani \&
Shibata (2001) have recently shown that the precise location of the
outer gap will itself depend on the inflow and outflow of current,
suggesting further non-steady feedback processes. The implication of
these exercises is that we ignore interactions between the polar cap
and outer gap at our peril.

\section*{Conclusions}
This paper represents an attempt to take a novel view of the pulsar
phenomenon. By abandoning the view that a pulsar's
magnetosphere is in a near-steady state, and further that its
behaviour on all scales of time and space is determined by so far
unexplained, yet complex events on the tiny polar caps, it is argued
that a promising new approach is possible.

Thus the magnetosphere is seen as having an inclined, essentially
dipolar, structure, whose apparently complex emission arises not from
abitrarily complex magnetic field components, but from subtle time-delayed
interactions between regions relatively remote from one another. The
principle interaction exists between the magnetic polar regions of the
neutron star and the outer gap. But it is also arguable that, quite
possibly, a similar aurora-like mirror interaction occurs between the
poles---a feature which might be mathematically represented as a
particle ``pressure'' exerted within or at the surface of the closed
corotating ``dead'' zone (Mestel \etal\ 2003). Each region of the
magnetosphere is then dependent on every other, yet never in a steady
state and always with a natural time lag. This idea is distinctly different 
from the more conventional view that the flow is driven smoothly from 
the tiny polar regions to the light cylinder, and opens a whole new line 
of pulsar investigation.

In developing our specific ideas we lean on both the observational 
analyses of one of us (ETI-VIII) and the recently developed concepts of 
the other (Wright 2003a), where the geometry of the feedback process 
has been examined in greater detail and successfully compared with 
observations in a few highly-organised pulsars. Such a feedback system 
can naturally proceed to bifurcations (\ie, alternate states) and ultimately 
to fully chaotic emission without any need to invoke strange geometries 
or external influences.  This behaviour is highly reminiscent of what is 
found in real pulsars, and thereby hints at the possibility of uniting theory 
and observations.

Here an attempt is made to link these new ideas to the principal 
long-standing conundrums found in the study of older, slow pulsars. 
But the wider purpose is to suggest that future theoretical investigations
may benefit from links to existing studies of the properties of time-dependent 
non-linear systems, which demonstrate that highly complex behaviour 
can be found in even the simplest systems. Similarly, the subtle statistics 
of time series, often used to mine information from apparent chaos in fields
far removed from astrophysics, might be usefully applied to pulsar sequence data.
It would not be the first time that cross-discipline studies have given 
unexpected insights.

\acknowledgements
One of us (JMR) wishes to acknowledge the support both of the US
National Science Foundation Grant AST 99-87654 and of a visitor grant
from the Nederlandse Organisatie voor Wetenschappelijk Onderzoek. The 
other (GAEW) thanks the University of Sussex for a Visiting Research 
Fellowship, and also the University of Vermont for support from the 
above-mentioned US National Science Foundation Grant.

\end{document}